\documentclass[trackchanges]{aastex7}

\usepackage{amsmath}

\usepackage{placeins}
\usepackage{subcaption} 
\usepackage{csquotes}
\MakeOuterQuote{"}

\begin{document}

\title{Probing the Nature of Dark Matter Self-Interactions Through Observations of Massive Black Hole Mergers}
\shorttitle{Probing the Nature of Dark Matter Self-Interactions}
\shortauthors{Hoelscher et al.}

\correspondingauthor{Zachary J. Hoelscher}
\author{Zachary J. Hoelscher}
\affiliation{Department of Physics and Astronomy, Vanderbilt University, Nashville, TN 37235, USA}
\email[show]{zachary.j.hoelscher@vanderbilt.edu}  

\author{Kelly Holley-Bockelmann}
\affiliation{Department of Physics and Astronomy, Vanderbilt University, Nashville, TN 37235, USA}
\affiliation{Department of Life and Physical Sciences, Fisk University, Nashville, TN 37208, USA}
\email[show]{k.holley@vanderbilt.edu}  

\author{Akaxia Cruz}
\affiliation{Center for Computational Astrophysics, Flatiron Institute, 162 5th Avenue, New York, NY 10010, USA}
\affiliation{Department of Physics, Princeton University, Princeton, NJ 08544, USA}
\email[show]{akaxia@princeton.edu} 

\author{N. Nicole Sanchez}
\affiliation{The Observatories of the Carnegie Institution for Science, 813 Santa Barbara Street, Pasadena, CA 91101, USA}
\affiliation{Cahill Center for Astronomy and Astrophysics, California Institute of Technology, MC249-17, Pasadena, CA 91125, USA}
\email[show]{nsanchez@carnegiescience.edu}

\begin{abstract}
Though the nature of dark matter remains elusive, two models have come to prominence with testable predictions: cold dark matter (CDM) and self-interacting dark matter (SIDM). While CDM remains the widely accepted model, SIDM was introduced to potentially help resolve the discrepancies between the predictions of the CDM model and observational data, in particular the predicted central density profiles. Previous work involving simulations of small numbers of Milky Way-mass galaxies shows that SIDM with a constant cross section of 1 $\rm{cm^2/g}$ delays massive black hole (MBH) mergers as compared to CDM when the host halo has a flattened central density profile. In this work, we use mock gravitational wave observations of MBH mergers to test LISA's capability to indirectly probe dark matter physics. As a proof of concept, we use zoom-in simulations of two galaxy evolutionary histories to show that LISA may be able to distinguish (with a p--value $\leq$ 0.05) between CDM and SIDM with a short-range interaction and a constant cross section of 1 $\rm{cm^2/g}$, provided at least $\sim70$ MBH mergers are observed with signal-to-noise ratios greater than 10. Given our small sample size, this should be regarded as illustrative, rather than definitive. We emphasize that our work does not consider more realistic models with a velocity-dependent cross section, though our exploratory work shows that LISA may provide a pathway to probe dark matter self-interactions, motivating future work with more realistic, currently-favored models and larger simulation suites. 
\end{abstract}

\keywords{\uat{Black Hole Physics}{159} --- \uat{Dark Matter}{353} --- \uat{Galaxy Dark Matter Halos}{1880} --- \uat{Cosmology}{343}}


\section{Introduction} 
\label{sec:intro} 

Dark matter makes up around 85 percent of the mass of the Universe, though its nature remains a mystery~\citep[see][and references therein for a review]{Review}. Since the first convincing evidence for dark matter was seen in galaxy clusters~\citep{Zwicky_1933, Zwicky_1937} and galaxy rotation curves~\citep{2}, the evidence for its existence has only grown. Cold dark matter (CDM) has been successful in explaining a wide span of observations, from the power spectrum of the cosmic microwave background~\citep{Planck_2018}, to the imprint of baryonic acoustic oscillations on the large-scale structure of galaxies~\citep{DESI_1, DESI_2}, to observations of galaxy clusters ~\citep{Clowe_2006, Mandelbaum_2006, Umetsu_2016}.

While CDM has many successes, there remain astrophysical challenges. Simulations of CDM without baryons show halos that follow a cuspy Navarro-Frenk-White (NFW) density profile ~\citep{Dubinski_1991, 11, Navarro_1997, Ludlow_2013}, whereas dwarf galaxies are inferred to have a broad range of density profiles, some of which are cored~\citep{6, Cored_Profile_Galaxies, Oman_2015, Santos_2017, Santos_2020}. The variation of central density profiles for dwarf galaxies in the same mass range is related to the "diversity of rotation curves" problem~\citep{Oman_2015, Kamada_2017}. This is closely related to the "core-cusp controversy" which refers to a tension between expected cuspy profiles and observed cored profiles in some dwarf galaxies. While it is known that baryonic feedback can produce cored profiles through the effects of supernovae outflows, it may be difficult to explain the observed cores through these means~\citep{Oman_2015, Santos_2020}, but see \cite{Cruz2025}. 

While there are possible challenges to CDM that could support alternative dark matter models, we also note that there is a large body of work demonstrating that astrophysical processes could possibly transform CDM halos to match observations. Supernova heating and energetic feedback from accreting black holes could alter a CDM-only NFW halo to create the cored halos we observe today~\citep{Cores_Dwarf_Galaxies, Read_Gilmore_2005, Pontzen_2012,
Governato_et_al_2012, Teyssier_2013, Cintio_2013, Onorbe_2015, Chan_2015, Tollet_2016, Read_2016, Dutton_et_al_2019, 
Cores_Dwarf_Galaxies2, Lazar_2020, Jahn_2023, Azartash_2024}. However, stellar feedback as the global solution to the core-cusp problem may be challenged by the discovery of dwarf galaxies with both cored density profiles and far too few stars to support core formation via feedback~\citep{Baryonic_Feedback_Not_Enough}, though this may be model dependent. Alternatively, galaxy mergers could form a cored density profile, although it seems that galaxy mergers are not enough to produce the observed cores~\citep{Merger_Core}. Another potential solution involves flattening the central density profile due to torques placed on the dark matter halo by a rotating galactic bar~\citep{BarFlattensHalo}, or scouring by few-body scattering from a massive black hole binary~\citep[e.g]{BBHcores}.

A promising alternative to CDM is self-interacting dark matter (SIDM)~\citep{6}. Unlike standard CDM~\citep{5}, this class of models consists of particles that have non-gravitational interactions with one another, resulting in elastic collisions of particles~\citep{6}. Since SIDM particles are assumed to be non-relativistic, this is then a modification to the CDM model in which non-gravitational interactions between particles are possible.

The major advantage of SIDM is that it can naturally produce a flattened central density profile in dwarf galaxies, which alleviates some of the tensions inherent in the CDM-only model~\citep{6}. Even more promising, SIDM can eventually produce cuspy profiles via core collapse, if the halo is allowed to evolve for a long enough time~\citep{Hennawi_2002, Essig_2019, Shah_2024, Zeng_2025}, also see~\citep{Ostriker_2000}. This variation of SIDM halo density profiles may result in a diversity of rotation curves~\citep{Kamada_2017, Ren_2019}.  

A recent study of galaxy clusters constrains the interaction cross section to $\sigma/m < 0.19$ $ \rm{cm^2~g^{-1}}$, which may be too low to produce cored density profiles in dwarf galaxies -- unless the cross section varies with velocity~\citep{SIDM_Cross_Section_Bound}, which other works support~\citep{Sagunski_2021, Andrade_2021, Gopika_2023}, also see~\citep{Annika_Peter_2013, Kim_2017}. A velocity-dependent SIDM model is predicted from a Yukawa interaction with a light mediator, and such a model may satisfy observational constraints~\citep{Sagunski_2021, Correa_2021, Correa_2022}. 

The different central density profiles produced by SIDM and CDM with baryons can manifest in larger changes in the way a galaxy evolves. There are also impacts on massive black hole dynamics, with SIDM resulting in more off-center black holes in low-mass galaxies at late times~\citep{DiCintio_2017}. ~\cite{13} uses high-resolution hydrodynamic N-body zoom-in simulations of Milky Way-mass galaxies to show that massive black hole inspiral and merger timescales are delayed in SIDM with a constant cross section of 1 $\rm{cm^2 g^{-1}}$. (For brevity, we hereafter refer to SIDM with a constant cross section of 1 $\rm{cm^2 g^{-1}}$ as SIDM-1.) The delay in black hole growth via the merger channel results in less black hole accretion, which triggers less AGN feedback that would otherwise have inhibited star formation at the Milky Way-mass scale. They also show that over the lifespan of a galaxy at the Milky Way-mass scale, when embedded in a SIDM-1 halo, about three times more stars form than in a CDM halo. Finally, they show that by the present day, the central black holes in both SIDM-1 and CDM halos have similar masses, so the nature of dark matter may not be encoded in local black hole demographics. 
To explore how well CDM and SIDM-1 could be differentiated via gravitational wave observations, we study trends in massive black hole merger histories as a proof of concept.

In this paper, we make predictions from galaxy simulations with CDM and SIDM-1 to test the observability of the effects of dark matter self-interactions on the assembly of massive black holes. Recent updates to SIDM work prefer a velocity-dependent cross section, which we plan to include in a future work. The constant cross section used here, however, is a useful approximation at the Milky Way scale, and in rough agreement with current constraints~\citep{Correa_2021, Correa_2022} at this scale. We also note that a velocity-dependent cross section can be approximately represented by a constant effective cross section of $\mathcal{O}(1)$ cm$^2$/g at the Milky Way scale~\citep{Yang_2022}. The velocity dependence of the scattering cross section is primarily important when one expects to have a large velocity range, which would occur if the simulations included galaxies with a wide range of halo masses. For example, one would need to consider velocity-dependence of the scattering cross section if the simulations included low-mass galaxies. In future work, which must incorporate a much larger suite of halos with a variety of masses, this will be important, though it is not a major concern here, as we strictly consider galaxies at approximately the Milky Way mass scale. For the present work, we thus do not expect velocity dependence of the scattering cross section to have a large effect. While the constant cross section incorporated here can be considered as a reasonable approximation, the reader should still note that this work is not directly applicable to SIDM with a velocity-dependent cross section. Our results should be viewed as an intriguing proof-of-concept, rather than definitive.

Here, we model gravitational wave observations of massive black hole mergers as detected by LISA (Laser Interferometer Space Antenna). LISA is a space-based gravitational wave observatory set to launch in 2035, as a joint project of NASA and the European Space Agency. LISA will consist of a constellation of three spacecraft arranged in a triangular formation to form an interferometer, which will allow the detection of gravitational waves with frequencies between around 0.1 mHz and 1.0 Hz~\citep[for a review, see][]{colpi2024lisa}. One of the primary science objectives of LISA is to detect black hole mergers between $10^5$ and $10^7$  M$_\odot$ out to redshifts greater than 10; this makes LISA an ideal tool to track the merger history of massive black holes over cosmic time~\citep{Colpi19}.
 
The goal of this paper is to determine whether, over a four-year LISA mission, it would be possible to discriminate between SIDM-1 and CDM through the impact of dark matter self-interactions on the cosmic times and mass ratios of massive black hole mergers. We note that there is potential in using phase shifts of gravitational waves to investigate the nature of SIDM, which is a related, but distinct method~\citep{Banik_2025}. 

The paper is organized as follows: in Section \ref{sec:SimulationParameters} we describe the simulations and provide cumulative distribution functions (CDFs) for cosmic times and mass ratios of massive black hole mergers. In Section \ref{sec:Results}, 
we describe our procedure for discriminating between CDM and SIDM-1 and discuss our implementation of the Kolmogorov-Smirnov (KS) test. We also provide estimates of the discriminating power of merger events with respect to CDM and SIDM-1 models. We then briefly comment on the signal-to-noise ratios (SNR) for the mergers, as they would be seen in LISA. In Section~\ref{sec:Conclusions} we summarize our conclusions and provide recommendations for future work. 

\section{Simulation Parameters} \label{sec:SimulationParameters}

We analyze zoom-in simulations of Milky Way-mass galaxies selected from the Romulus25 volume~\citep{Parameters} using the Charm N-body GrAvity Solver (ChaNGa) code~\citep{CHANGA}. We focus on two galaxies, \enquote{GM4} and \enquote{GM7}, which are of similar mass and are produced through a genetic modification process, which subtly perturbs the initial overdensity of a fiduical galaxy to generate a final suite of galaxies with slightly different assembly histories~\citep{PontzenGM}. We run each with both CDM and SIDM-1, thus we have four distinct simulations. The simulations have a final central SMBH mass of $\mathcal{O}(10^7)$ M$_{\odot}$. In Table~\ref{table:MassTable}, we provide some details on the galaxies simulated, where the masses shown are for redshift $z \approx 0$. 

\FloatBarrier 
\begin{table}[h!]
  \centering
    \begin{tabular}{p{2 cm} p{2 cm} p{2 cm} p{2 cm} p{3cm} p{4cm}}
    \hline
    Galaxy & Total Mass & Stellar Mass & Gas Mass & Dark Matter Mass & Central SMBH Mass\\
    \hline
    GM4 CDM & $7.24 \times 10^{11}$ & $1.04 \times 10^{10}$ & $6.36 \times 10^{10}$ & $6.50 \times 10^{11}$ & $6.75 \times 10^{7}$\\
    GM7 CDM & $7.85 \times 10^{11}$ & $1.25 \times 10^{10}$ & $7.41 \times 10^{10}$ & $6.99 \times 10^{11}$ & $5.85 \times 10^{7}$\\
    GM4 SIDM-1 & $8.73 \times 10^{11}$ & $3.89 \times 10^{10}$ & $9.50 \times 10^{10}$ & $7.39 \times 10^{11}$ & $4.41 \times 10^{7}$\\
    GM7 SIDM-1 & $9.80 \times 10^{11}$ & $2.07 \times 10^{10}$ & $1.10 \times 10^{11}$ & $8.50 \times 10^{11}$ & $4.75 \times 10^{7}$\\
    \hline
    \end{tabular}
  \caption{Here we provide the total mass, stellar mass, gas mass, dark matter mass, and central SMBH mass for each of our simulated galaxies at $z \approx 0$, where the units are solar masses.}
  \label{table:MassTable}
\end{table}
\FloatBarrier 

ChaNGa is an updated form of GASOLINE, a smoothed-particle hydrodynamics code~\citep{Wadsley_2017}, that is designed to scale efficiently at the 100,000-core level, making it an effective tool for tackling computationally intensive problems in theoretical astrophysics~\citep{CHANGA}.  ChaNGa and GASOLINE have been used extensively to study ultra-diffuse galaxy formation~\citep{UltraDiffuseGalaxies}, quenching of star formation~\citep{Sanchez2021}, ultra-diffuse dwarf galaxy properties~\citep{Applebaum2021}, and the relationship between stellar mass and halo mass for dwarf galaxies~\citep{Munshi_2021}. ChaNGa and GASOLINE are also used to study massive black hole formation~\citep{Bellovary2016, Dunn_2018, Dunn_2020}, SMBH-IMBH mergers~\citep{IMBH_Inspiral}, and the nature of the dark matter particle~\citep{Fry_2015, Cintio_2019, 13}. 

ChaNGa employs star formation models that include a Kroupa initial mass function (IMF) to describe the stellar mass distribution. It also incorporates models for blastwave supernovae feedback and metal line cooling, like in GASOLINE~\citep{Wadsley_2004, Blastwave, Metal_Diffusion, Sanchez_2019}. The code approximates the effects of supernovae feedback by temporarily disabling radiative cooling in gas particles near young stars, allowing the simulation to incorporate heating that would otherwise be unresolved due to resolution limitations~\citep{Blastwave}. ChaNGa also incorporates subgrid models to more accurately apply dynamical friction to massive black hole dynamics~\citep{Tremmel_Dynamical_Friction}. 

Similar Milky Way-mass galaxy simulations were introduced in~\citep{Sanchez_2019} for CDM and~\citep{13} for SIDM-1. The initial conditions are selected from a dark matter-only volume that is 50 Mpc on each side, and is resimulated at a higher resolution with baryonic physics using the method of~\cite{Zoom_In}. 
The initial conditions for the simulation suite are produced using genetic modification~\citep{Roth2015}, resulting in two Milky Way-mass galaxies at redshift zero with a dark matter virial mass of $\sim 7 \times 10^{11}$ M$_\odot$.
The simulations have a softening length of 250 pc, a mass resolution of $1.4 \times 10^5$ M$_\odot$ for dark matter, and $2.1 \times 10^5$ M$_\odot$ for gas particles. The simulations assume a $\Lambda$-dominated cosmology, with $\Omega_m=0.3086$, $\Omega_{\Lambda} = 0.6914$, $h=0.67$, and $\sigma_8 = 0.8288$~\citep{Planck_2014}. 

\begin{figure}[h!]
    \centering
    \begin{subfigure}[b]{0.48\textwidth}
        \centering
        \includegraphics[width=3.5in, height=2.5in]{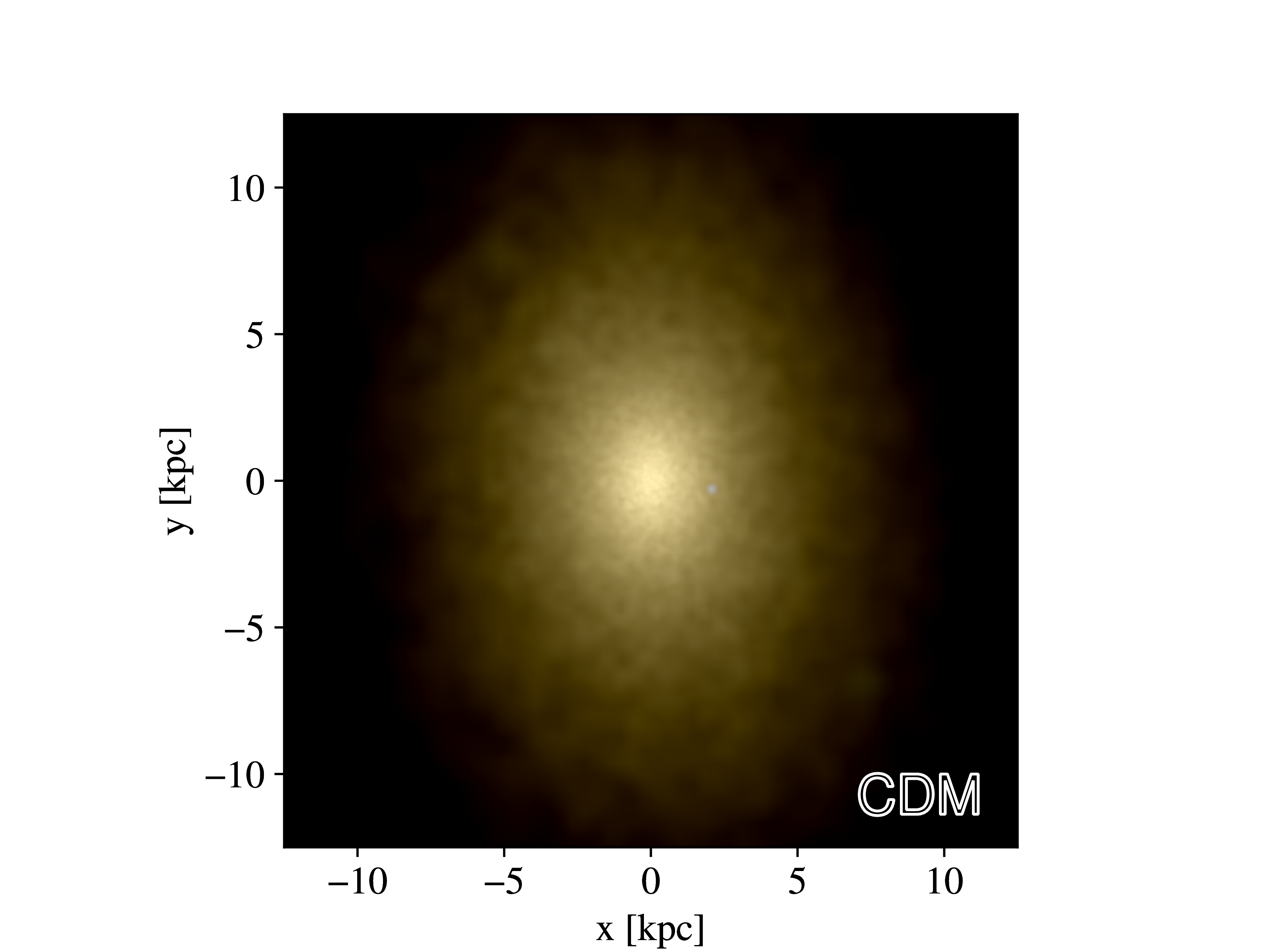}
        \label{fig:GM4_CDM_labeled}
    \end{subfigure}
    \hfill
    \begin{subfigure}[b]{0.48\textwidth}
        \centering
        \includegraphics[width=3.5in, height=2.5in]{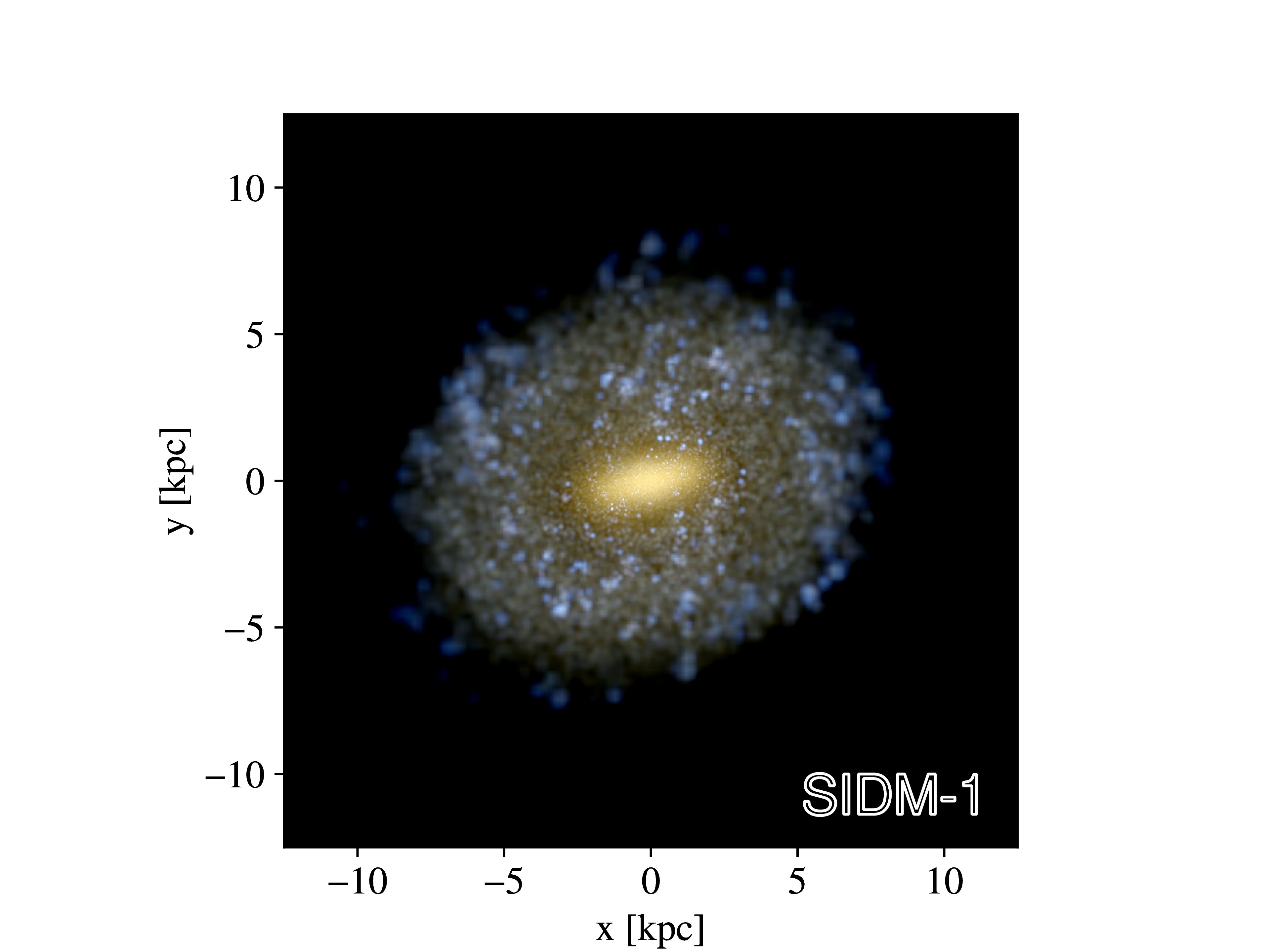}
        \label{fig:GM4_SIDM_labeled}
    \end{subfigure}
   \caption{One of the simulated galaxies with CDM (left) and SIDM-1 (right), at redshift zero. With CDM, the total mass (gas + stars + dark matter) is $7.24 \times 10^{11}$ M$_{\odot}$, the stellar mass is $1.04 \times 10^{10}$ M$_{\odot}$, and the dark matter mass is $6.50 \times 10^{11}$ M$_{\odot}$. With SIDM-1, the total mass is $8.73 \times 10^{11}$ M$_{\odot}$, the stellar mass is $3.89 \times 10^{10}$ M$_{\odot}$, and the dark matter mass is $7.39 \times 10^{11}$ M$_{\odot}$. These masses are calculated using Pynbody, and include all particles that are gravitationally bound to the halo~\citep{Pynbody}. Pynbody uses a stellar population model to estimate the stellar luminosity in different wavelength bands, from which it determines the apparent colors in the images. The colors are set by mapping red, green, and blue to the I (infrared), V (visible), and U (ultraviolet) bands, respectively. Here, the clearly visible differences are primarily due to the effects of SIDM-1 on the star-formation history. As SIDM-1 delays the growth of black holes via mergers, star formation is not suppressed to the same degree.}
    \label{fig:GM4_Galaxy_Image}
\end{figure}

Similar simulations have previously been used to test the impact of AGN feedback on the circumgalactic medium~\citep{Sanchez_2019}, quenching in Milky Way-mass galaxies~\citep{Sanchez2021}, and massive black hole formation in Milky Way-mass galaxies with CDM halos and SIDM-1 halos~\citep{13}. We provide representative images (Figure~\ref{fig:GM4_Galaxy_Image}) of one of the galaxies used in this work. Greater black hole accretion causes star formation to be suppressed in the galaxy evolved in a CDM halo.

\FloatBarrier

\FloatBarrier

\subsection{Self-Interacting Dark Matter}
In general, the SIDM interaction rate $\Gamma_{\rm SI}$, as a function of position, $r$, and redshift, $z$, is expressed as follows~\citep{SIDM, Fry_2015, 13}.
\begin{equation}
    \Gamma_{\rm SI}(r,z) \propto \rho_{\rm DM}(r,z) v_{\rm DM}(r,z) \sigma_{\rm DM}
\end{equation}

\noindent In the equation above, $\rho_{\rm DM}$ is the dark matter density, $v(r,z)$ is the velocity dispersion, and $\sigma_{\rm DM}$ is the dark matter cross section. The interaction probability is given by $\Gamma_{\rm SI} \delta t$, where $\delta t$ is the timestep~\citep{SIDM}. In the simulations, we allow particles to scatter when 
a uniform random variable between 0 and 1 is less than half the interaction probability.~\footnote{The factor of $1/2$ enters because the neighbor search is two sided~\citep{Fry_2015}. This means that we consider particle $A$ scattering with particle $B$, as well as $B$ scattering with $A$, so a factor of $1/2$ is needed to prevent double-counting.} The density and relative velocities of dark matter particles are calculated by smoothing over the nearest 32 particles, and particles are allowed to scatter at random angles~\citep{Fry_2015}. The interactions between particles are modeled as elastic collisions that cause particles to isotropically scatter and exchange energy. For the cross-section and mass scale considered in this work, this leads dense regions, such as the centers of dark matter halos, to become isothermal, producing cores~\citep{Burkert_2000, Balberg_2002, Ahn_2005, Koda_2011, Fry_2015, Elbert_2015, Zeng_2022, Tran_2024}. Eventually, the core can collapse, so the profile evolves to become steeper than NFW, though this is dependent on the mass of the DM halo and its concentration~\citep{Burkert_2000, Balberg_2002, Koda_2011, Turner_2021, Zeng_2022, Shah_2024, Tran_2024, Zeng_2025, Tran_2025}. 

\subsection{Black Hole Physics}
Massive black hole seeds of $10^6$ M$_\odot$ are allowed to form once gas particles have a density of $\sim 45$ amu $\rm{cm^{-3}}$, a temperature of 9,500 to 10,000 Kelvin, and a low metallicity $Z < 3 \times 10^{-4}$. The particles are also required to meet the conditions necessary for star formation. These conditions result in massive black hole seeds forming in regions of high density, or where gas cools slowly~\citep{MergeConditions}. The majority of such seeds are produced during the first 1 Gyr~\citep{Parameters, 13}. 
Such formation times are similar to what is expected to have occurred in our Universe~\citep[e.g.][]{Maiolino24,Priya24}. 

Dynamical friction~\citep{Chandra_Dynamical_Friction} causes massive black holes to migrate towards the center of a galaxy, where they can eventually merge~\citep{BBR80}. Some cosmological simulations underestimate dynamical friction due to resolution effects that fail to properly capture the short-range interactions; the simulations used in this work compensate for this effect by applying a drag force to the massive black holes that is consistent with dynamical friction generated by the nearby particles. The correction term to the SMBH acceleration, $\textbf{a}_{\rm DF}$, takes the form below~\citep{Tremmel_Dynamical_Friction}: 

\begin{equation}
    \textbf{a}_{\rm DF} = -4 \pi G^2 M \rho(< v_{\rm BH})\ln(\Lambda) \frac{\textbf{v}_{\rm BH}}{v_{\rm BH}^3}.
\end{equation}

\noindent Here $\textbf{v}_{\rm BH}$ is the velocity of the black hole, $M$ is the black hole mass, $\rho(<v_{BH})$ represents the density of particles whose speeds are lower than that of the black hole, and $\ln(\Lambda)$ is the Coulomb logarithm. The Coulomb logarithm can be given by: $\ln(\Lambda) \approx \ln(b_{\rm max}/b_{\rm min})$, where $b_{\rm max}$ is the softening length, $\epsilon_g$, and $b_{\rm min}$ is either the minimum radius for a 90 degree deflection ($GM_{BH}/v_{BH}^2$), or the Schwarzschild radius, whichever is greater~\citep{Tremmel_Dynamical_Friction}. 

The simulations incorporate an accretion model that modifies Bondi-Hoyle accretion to take into account rotational support from the gas surrounding the black holes~\citep{BondiHoyle1, BondiHoyle2, BondiHoyle3, BoostedBondiHoyle, Parameters}. We provide this model below. Note that when calculating local density and temperature values, we smooth over the closest 32 gas particles~\citep{Parameters}. 

\begin{equation}
    \dot{M}_{BH} = \alpha (n) \pi (G M_{BH})^2 \rho f(v_{\rm bulk})
\end{equation}

\noindent Here $\rho$ refers to the local gas density, $v_{\rm bulk}$ is the bulk velocity of the gas, and $M_{BH}$ is the black hole mass. We boost the accretion rate by a density-dependent factor $\alpha$, which prevents low resolution from causing an underestimate of the accretion rate~\citep{BoostedBondiHoyle, Parameters}:

\begin{equation}
    \alpha (n) = 
    \left\{
        \begin{array}{lr}
            ( \frac{n}{n_*} )^2, & \text{if } n \geq n_* \\
            1, & \text{if } n < n_* ,\\
        \end{array}
    \right.
\end{equation}

\noindent where $n$ is the number density of gas particles and $n_*$ is the star formation threshold. We use $f(v_{\rm bulk})$ to describe the function below, which is used to correct Bondi-Hoyle accretion to account for rotational support.

\begin{equation}
    f(v_{\rm bulk}) = 
    \left\{
        \begin{array}{lr}
            1/(v^2_{\rm bulk} + c_s^2)^{3/2}, & \text{if } v_{\rm bulk} > v_{\theta} \\
            c_s/(v_{\theta}^2 + c_s^2)^2, & \text{if } v_{\rm bulk} < v_{\theta}\\
        \end{array}
    \right.
\end{equation}

\noindent Here $v_{\theta}$ represents the tangential velocity, $v_{\rm bulk}$ the bulk velocity of the gas, and  $c_s$ the speed of sound in the gas. Energy from accretion is allowed to isotropically transfer to the gas surrounding the black holes. The rate is given by the equation below, where $\epsilon_r$ is the radiative efficiency, $\epsilon_f$ is the feedback efficiency, $c$ is the speed of light, and $\dot{M}_{BH}$ is the accretion rate ~\citep{Parameters}: 

\begin{equation}
    \dot{E}_{BH} = \epsilon_r \epsilon_f \dot{M}_{BH} c^2.
\end{equation}

\noindent Note that $\epsilon_r = 0.1$ and $\epsilon_f = 0.02$ ~\citep{Parameters, Sanchez_2019, 13}. The energy transferred from the accretion disk to the surroundings can generate galactic-scale outflows that can enrich the circumgalactic medium and quench star formation without evacuating the gas halo~\citep{Quench, RomulusC, Sanchez_2019, Garza_2024}. 

A complete discussion of black hole physics must address the conditions that lead to mergers. We consider two massive black holes to have merged once their separation is less than two softening lengths and their relative velocities are small enough that they would be gravitationally bound. These conditions are equivalent to satisfying the equation below, where $\Delta \textbf{r}$, $\Delta \textbf{v}$, and $\Delta \textbf{a}$ are the vectors for relative separation, velocity, and acceleration, respectively~\citep{MergeConditions, Parameters}: 

\begin{equation}
    \frac{1}{2} \Delta \textbf{v}^2 < \Delta \textbf{a} \cdot \Delta \textbf{r}.
\end{equation}

\noindent These conditions simplify the simulations by avoiding detailed modeling of the final inspiral, but leave out some important physics. A full treatment would consider the effects from gravitational wave emission, few-body scattering, and accretion disk mediated inspiral, as all of these can work to force black holes to merge. Such effects are important to determine how long it takes a binary to merge, which is an active area of research~\citep[e.g]{Finalpc}. It is, however, impractical to model these in full detail, given the resolution limitations of the simulations, though future studies could incorporate simple estimates derived from recent work, e.g.~\citep{Holley-Bockelmann_2025}. Such effects could delay merger times relative to what we see in our simulations, though since this would occur with both CDM and SIDM-1, it would not necessarily eliminate the delaying effect of SIDM-1 relative to CDM.

\subsection{Characterizing merger histories}

\noindent We begin by defining the following quantities:

\begin{enumerate}\setlength{\itemsep}{0pt}
    \item $\rm{N_{SIM}}$: Total number of mergers in the simulations 
    \item $\rm{N_{MOCK}}$: Number of mergers hypothetically observed by LISA; this is not a calculated quantity, but rather a parameter we vary
    \item $\rm{N_{DIFF}}$: Number of mergers needed to differentiate between CDM and SIDM-1 at a statistically significant level; this value depends on whether we consider cosmic times or mass ratios
\end{enumerate}
\noindent Note that the number of massive black hole mergers LISA will actually see is extremely uncertain~\citep{LISA_Detection_Rate_and_Review}, and may be smaller or larger than $\rm{N_{DIFF}}$.

As we have simulations of only two galaxies, our sample size is our biggest limitation, meaning our results should be regarded as illustrative values only. The simulations provide us with cosmic times and mass ratios for $\rm{N_{SIM}}$ massive black hole mergers. We combine the sets of the cosmic times and mass ratios of each MBH merger in our two simulated galaxies into one dataset for CDM, and one dataset for SIDM-1, so $\rm{N_{SIM}}$ is the sum of the number of mergers for the two simulated galaxies. Table~\ref{table:MergersTable} shows the number of massive black hole mergers in each simulated galaxy in our simulation suite;   $\rm{N_{SIM}} =51$ for CDM, and $\rm{N_{SIM}} =24$ for SIDM-1. 

\begin{figure}[h!]
    \centerline{
    \includegraphics[width=3.5in]{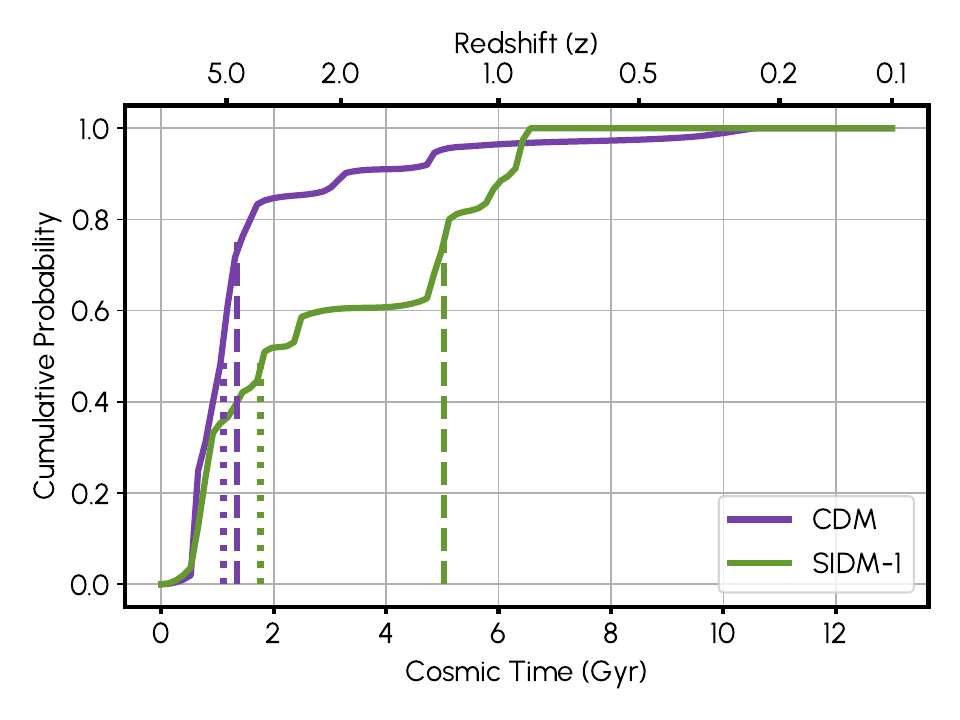} \includegraphics[width=3.5in]{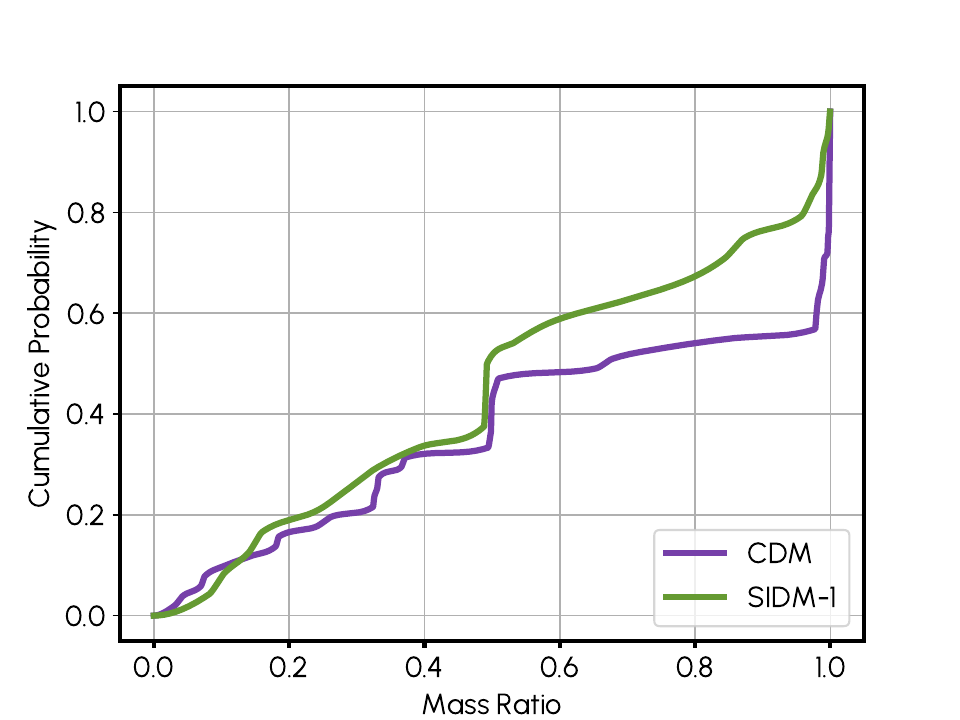}
    }
    \caption{Cumulative probability distributions of massive binary black hole merger properties for galaxies simulated with CDM (purple) and SIDM-1 (green). Left: Cumulative massive binary merger probability as a function of cosmic time. Note that the CDM model exhibits more rapid assembly of massive black holes by mergers at early times. We show the medians with doted lines, and third-quartiles with dashed lines. Right: Cumulative massive binary merger probability as a function of mass ratio. The SIDM-1 model favors moderate mass ratio mergers, while nearly half the mergers in the CDM model are nearly equal mass.}
    \label{fig:Both_CDFs}
\end{figure}

We then generate cosmic time and mass ratio cumulative distribution functions (CDF) from the CDM and SIDM-1 data, for a total of 4 CDFs. These are made continuous using monotonic splines~\citep{24}. These CDFs are then inverted to produce inverse CDFs. Figure~\ref{fig:Both_CDFs} shows CDFs built from the CDM and SIDM-1 simulations for mass ratio and cosmic time of the massive black hole mergers, which are produced using the $\rm{N_{SIM}}$ mergers from our simulations. One can readily see that mergers are delayed in the SIDM-1 model. 

Thereafter, we draw 10,000 samples of size $\rm{N_{MOCK}}$ from each inverse CDF, varying $\rm{N_{MOCK}}$, to generate 4 sets of 10,000 empirical PDFs for each $\rm{N_{MOCK}}$. (This procedure is commonly referred to as inverse transform sampling.) These are convolved with a Gaussian kernel to de-emphasize the precise step-wise shape of our small dataset and better represent an underlying smooth distribution. We then integrate the empirical PDFs to produce empirical cumulative distribution functions (CDFs) to conduct a Kolmogorov-Smirnov (KS) test. (See subsection \ref{KSTest}.) The code used in this analysis is publicly available~\citep{Hoelscher2025}.

\FloatBarrier

\FloatBarrier

\FloatBarrier 
\begin{table}[h!]
  \centering
    \begin{tabular}{p{2cm} p{2cm}}
    \hline
    Galaxy & MBH Mergers \\
    \hline
    GM4 CDM & 33\\
    GM7 CDM & 18 \\
    GM4 SIDM-1 & 9 \\
    GM7 SIDM-1 & 15 \\
    \hline
    \end{tabular}
  \caption{The number of MBH mergers in our simulations for each of the two galaxies, with CDM and SIDM-1. GM4 and GM7 refer to particular galaxies whose initial conditions are set via genetic modification. This results in galaxies with similar evolution but different satellite populations~\citep{13}. We combine GM4 and GM7 to make one dataset for CDM, and one for SIDM-1, so $\rm{N_{SIM}}$ for CDM is 51, whereas $\rm{N_{SIM}}$ for SIDM-1 is 24. We use the MBH merger times and mass ratios to produce PDFs, which allow us to estimate what LISA would see.}
  \label{table:MergersTable}
\end{table}
\FloatBarrier 

\newpage 

\section{Results}
\label{sec:Results}

We find that around 70 mergers are sufficient to ensure that the 95th percentile for the p--value is below the 0.05 significance threshold, when comparing cosmic times, or 160 mergers, when comparing mass ratios, though we again emphasize that these should be regarded as illustrative values only. These values should not be interpreted as quantitative requirements for LISA, given the limitations of our small sample size, as cosmic variance may substantially change the number of mergers required. We see fewer mergers are needed when comparing cosmic times because the cosmic times CDFs for CDM and SIDM-1 are more different from one another than the mass ratios CDFs for CDM and SIDM-1. This suggests that the cosmic times provide a more direct indicator for the evolution of the halo density profile than the mass ratios, with our limited sample. 

\FloatBarrier

\begin{figure}[h!]
    \centerline{
    \includegraphics[width=3.25in, height=2.75in]{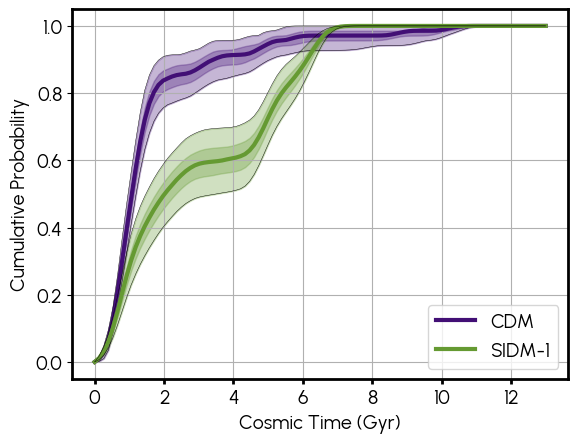} }
    \caption{We plot the distribution of 10,000 empirical CDFs for cosmic times of 70 MBH mergers, CDM (purple) and SIDM-1 (green).  The darker regions contain the 25th and 75th percentiles, whereas the lighter regions enclose the 5th and 95th percentiles. The dark line traces the 50th percentile. } 
    \label{fig:Wedge_Plot_Times}
\end{figure}

\begin{figure}[h!]
    \centerline{
    \includegraphics[width=3.25in, height=2.75in]{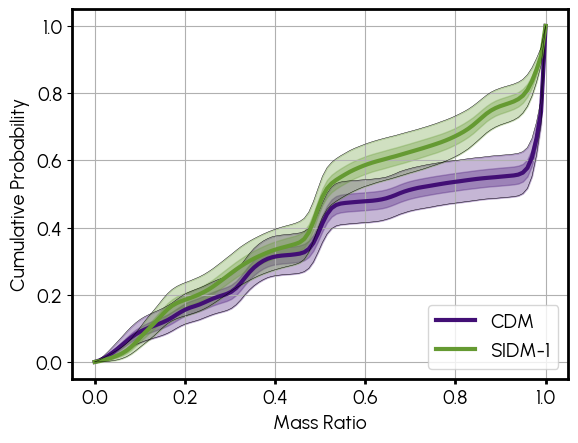} }
    \caption{We plot the distribution of 10,000 empirical CDFs for 160 MBH merger mass ratios, CDM (purple) and SIDM-1 (green).  The darker regions contain the 25th and 75th percentiles, whereas the lighter regions enclose the 5th and 95th percentiles. The dark line traces the 50th percentile. } 
    \label{fig:Wedge_Plot_Mass_Ratios}
\end{figure}

We include plots (Figure~\ref{fig:Wedge_Plot_Times} and Figure~\ref{fig:Wedge_Plot_Mass_Ratios}) to show how the 10,000 empirical CDFs are distributed. The darker region encloses the 25th and 75th percentiles, whereas the lighter region encloses the 5th and 95th percentiles. The dark line traces the 50th percentile. We show $\rm{N_{MOCK}} = 70$ for cosmic time, whereas $\rm{N_{MOCK}} = 160$ for mass ratios. 

\FloatBarrier

\FloatBarrier
\vspace{-0.75 cm}
\FloatBarrier

\FloatBarrier

\newpage

\subsection{Kolmogorov-Smirnov Test}
\label{KSTest}

To discriminate between CDM and SIDM-1 merger histories, we apply a KS test~\citep{19, Smirnov}, which compares the corresponding CDFs and yields a p--value quantifying the probability that two CDFs come from the same underlying probability distribution. A smaller p--value implies a more statistically significant difference between CDM and SIDM-1, allowing us to quantify the statistical significance of the difference between CDM and SIDM-1.

We compare CDM and SIDM-1 empirical CDFs for each of the 10,000 pairs of empirical CDFs. The application of the KS test allows us to compute 10,000 p--values. We can then see how the p--values are distributed. The KS test uses the maximum difference between each pair of empirical CDFs being compared, and the number of mergers ($\rm{N_{MOCK}}$) to compute p--values.  We vary $\rm{N_{MOCK}}$ to determine the number of mergers LISA would need to see over its lifespan to discriminate between CDM and SIDM-1 ($\rm{N_{DIFF}}$). 

The KS test is implemented as shown below~\citep{23} using equations~\eqref{eq:pvalue} and~\eqref{eq:altseries}. This gives the p--value as a function of $\rm{N_{MOCK}}$ and the maximum absolute difference (D) between the empirical CDFs being compared. 

\begin{equation}
    {\mathrm{p} \hspace{0.1 cm} \mathrm{value}} = \rm{Q_{KS}\bigg( \bigg( \sqrt{\frac{\rm{N_{MOCK}}}{2}} + 0.12 +\frac{0.11 \sqrt{2}}{\sqrt{\rm{N_{MOCK}}}} \bigg) D \bigg)}
    \label{eq:pvalue}
\end{equation}

\noindent To make the above expression more compact, we define the function $\rm{Q_{KS}}(x)$ as shown below. 

\begin{equation}
    \rm{Q_{KS}}(x) = \sum_{j=1}^{\infty} 2(-1)^{j-1} e^{-2j^2 x^2}
    \label{eq:altseries}
\end{equation}

This approximation for the p--value given in equation~\eqref{eq:pvalue} is valid for $\rm{N_{MOCK}} \geq 8$~\citep{23}. As equation~\eqref{eq:altseries} is an alternating series, the error from using a finite number of terms is bounded by the absolute value of the first omitted term. We cease adding terms when the absolute value of the first omitted term is less than $10^{-9}$, which ensures that the p--value is within $10^{-9}$ of the value one would get from using an infinite number of terms~\citep{MathMethodsTextBook}. 

\begin{figure}[h!]
    \centerline{
    \includegraphics[width=3in, height=2.4in, trim =0 0 0 1cm, clip=true]{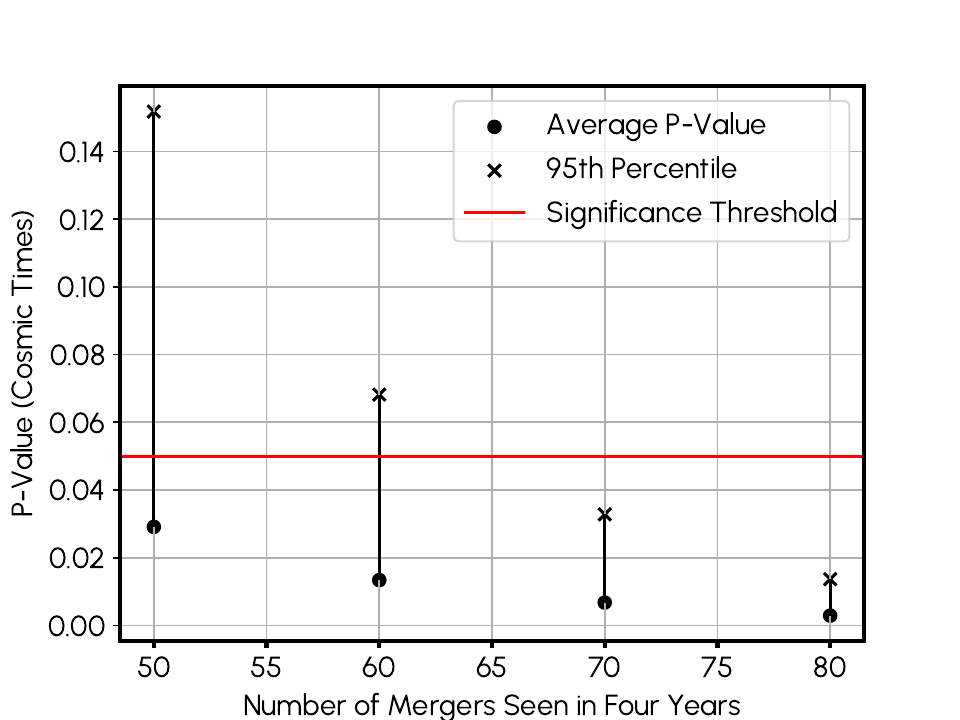} \includegraphics[width=3in, height=2.4in, trim =0 0 0 1cm,clip=true]{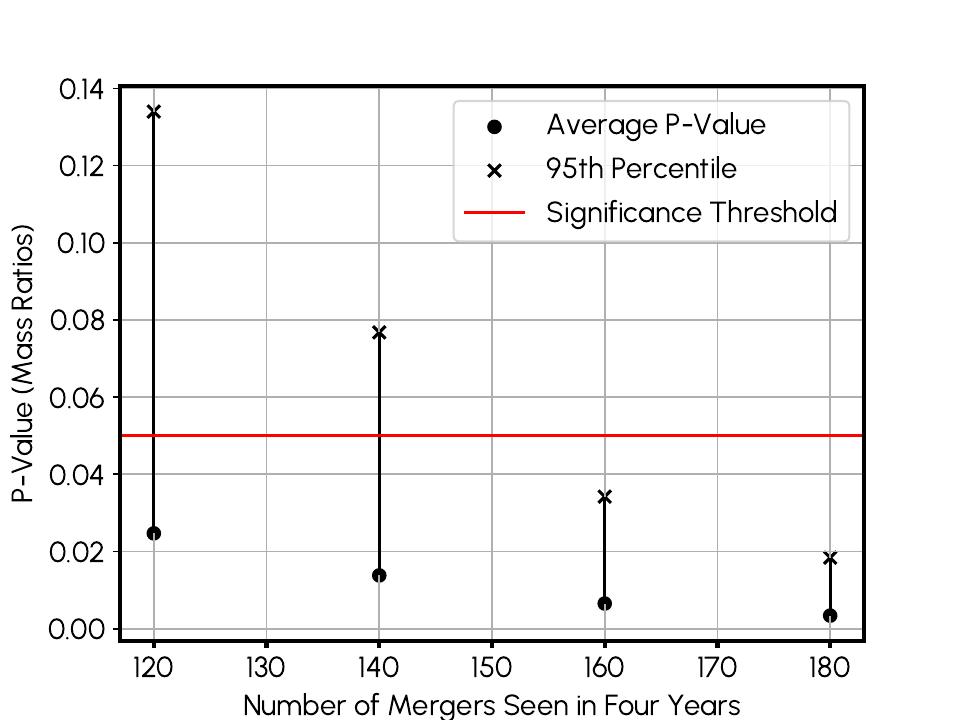}}
    \caption{Average p--values (black-dots) for comparing cosmic times (left) or mass-ratios (right) of mergers with CDM to mergers with SIDM-1. We indicate the 95th percentile with an X. The red line indicates the significance threshold of 0.05. For differentiating between CDM and SIDM-1, we wish to have the 95th percentile below the significance threshold. We find that $\sim70$ mergers are sufficient to attain this when comparing cosmic times, or $\sim160$ mergers when comparing mass ratios.}
    \label{fig:P_Value_95}
\end{figure}

\begin{figure}[h!]
    \centerline{
    \includegraphics[width=3.25in, height=2.75in]{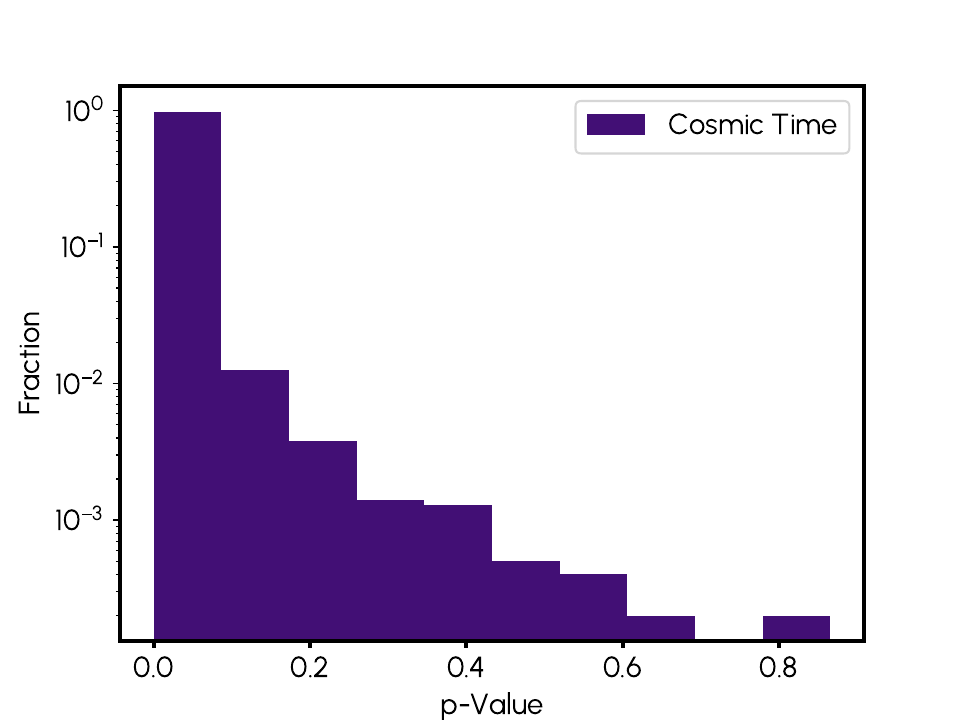} \includegraphics[width=3.25in, height=2.75in]{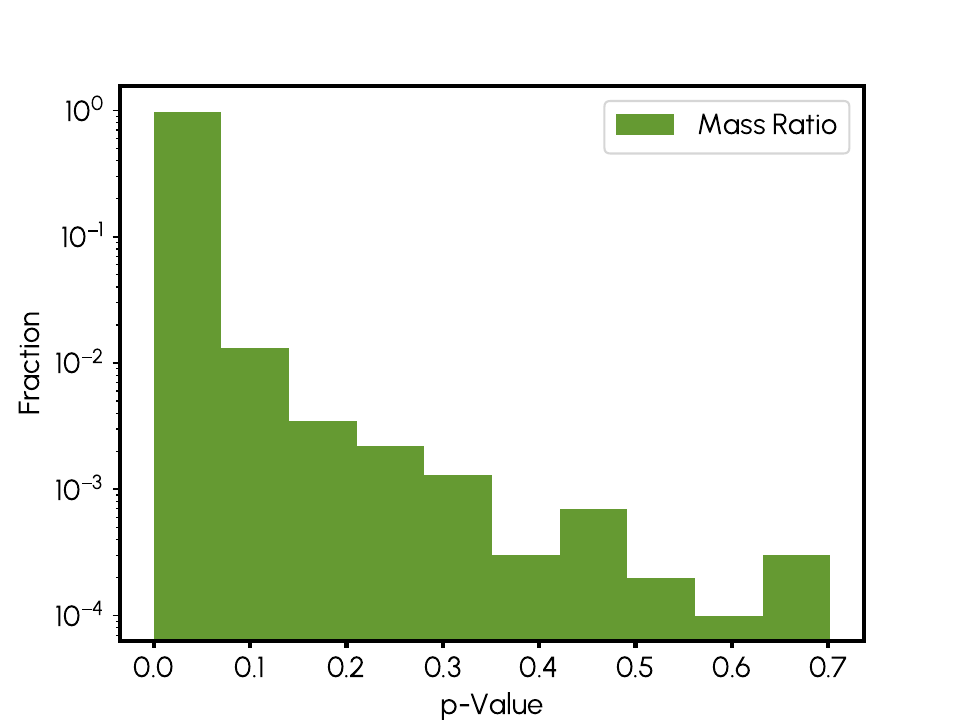}}
    \caption{Here we show the distribution of p-values for comparing cosmic times (left) and comparing mass ratios (right). For cosmic times, this corresponds to $\sim70$ mergers, whereas for mass ratios, this corresponds to $\sim160$ mergers. These plots show the fraction of the 10,000 p-values that live in each bin.} 
    \label{fig:P_Value_Distributions}
\end{figure}

We seek to determine the number of mergers needed to ensure that the 95th percentile for the p--value is below the 0.05 significance threshold. (This threshold implies that there is less than a five percent chance that the results are spurious.) We conduct this analysis twice; first comparing empirical CDFs for cosmic times, then comparing empirical CDFs for mass ratios. This allows us to determine different $\rm{N_{DIFF}}$ for each observable. We include plots below to show the average p--values and the 95th percentiles for cosmic times and mass ratios (Figure~\ref{fig:P_Value_95}). We also show the distributions for the p-values (Figure~\ref{fig:P_Value_Distributions}).

\vspace{-0.25 cm}

\FloatBarrier

\FloatBarrier

\FloatBarrier

\FloatBarrier

\subsection{Estimated Observability by LISA} 
\label{Observability}

The simulations provide cosmic times and mass ratios for massive black hole mergers, which are useful for comparing CDM and SIDM-1, though they do not provide black hole spin, orientation, sky localization, or orbital inclination. We can, however, estimate SNR for the GM7 set \footnote{We are missing some black hole mass information due to data loss for GM4, but given the similarity between the models and the very high SNR found in GM7, we are confident that all black hole mergers in the combined suite would be observable by LISA.} by assuming probability distributions for spins and inclinations and calculating the sky-averaged SNR.   We assume eccentricity is zero and that the spins are aligned. Although this should be considered a rough approximation, for our purposes, an order-of-magnitude estimate is sufficient.

\begin{figure}[h!]
    \centerline{
    \includegraphics[width=3.25in, height=2.75in]{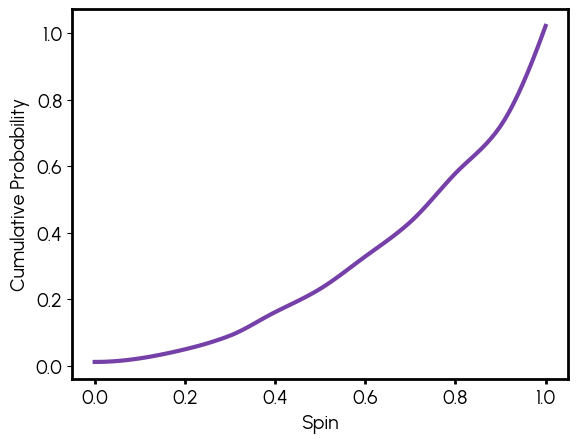}}
    \caption{Here we plot the cumulative distribution function corresponding to our assumed distribution for massive black hole spins.}
    \label{fig:Spin_Distribution}
\end{figure}

We assume that the cosine of the inclination is uniformly distributed between -1 and 1. We also assume that spins follow a distribution corresponding to the cumulative distribution function shown below (Figure~\ref{fig:Spin_Distribution}). We adopted this distribution from \citet{Ricarte_2025}, which inferred the spin population for massive black holes using X-ray reflection spectroscopy, while correcting for bias toward high spins due to the spin-dependence of radiative efficiency.~\footnote{We note that the function is approximately consistent with a beta distribution.} We generate 1,000 sample spins and inclinations from these distributions, and use IMRPhenomXHM~\citep{IMRPhenomXHM} to model the waveform of each sample merger. IMRPhenomXHM is an updated version of the waveform model IMRPhenomD~\citep{IMRPhenomD}, generating waveforms for binaries that have non-precessing spins and quasi-circular orbits. It is calibrated using a combination of post-Newtonian waveforms and waveforms produced via numerical relativity. We then calculate SNR for LISA using PyCBC, a gravitational wave Python package, assuming Planck14 cosmology~\citep{Planck_2014, AstroPy, PyCBC}, and incorporating a model for the projected LISA noise curve~\citep{Babak_2021}. 

\begin{figure}[h!]
    \centerline{
    \includegraphics[width=3.25in, height=2.75in]{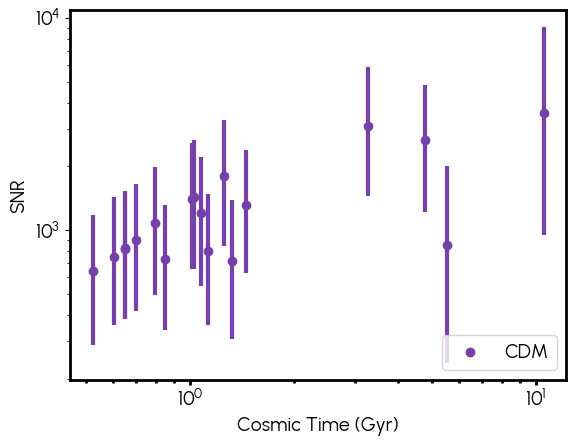} \includegraphics[width=3.25in, height=2.75in]
     {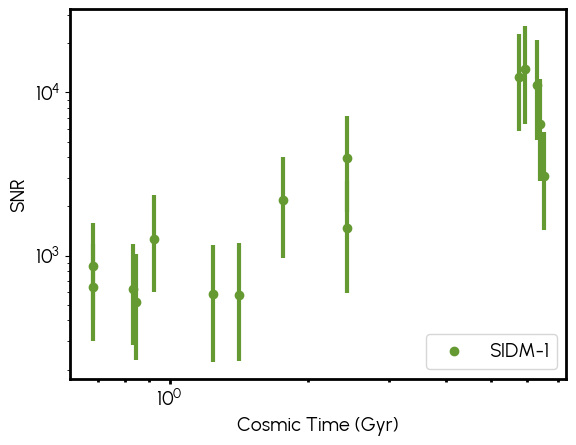}}
    \caption{Here we show the ranges of possible SNRs for each merger, by varying inclination and spin magnitude, for GM7 with CDM (left; purple) and SIDM-1 (right; green). Even with the most pessimistic SNR values, LISA would be able to detect such mergers, though note that these SNRs exclude the galactic foreground. We show the mean SNRs with dots. The error bars extend from the 5th percentiles to the 95th percentiles.}
    \label{fig:SNR_Plots}
\end{figure}

We note that our analysis does not account for the galactic binary foreground or confusion noise. While these SNR values could suggest that LISA would be able to observe all MBH mergers in Table~\ref{table:MergersTable} for galaxy GM7, this should not be regarded as exact, given the limitations and approximations discussed above. Below we show the assumed distribution of spins (Figure~\ref{fig:Spin_Distribution}), and the SNR ranges of our sample (Figure~\ref{fig:SNR_Plots}). 

\FloatBarrier

\FloatBarrier

\FloatBarrier

\FloatBarrier

\section{Conclusions and Recommendations for Future Work}\label{sec:Conclusions}

In this work, we employ two comparable Milky Way-mass galaxy simulation suites with differing dark matter models to determine whether LISA observations may be able to differentiate between them. We demonstrate, as a proof of concept, that LISA observations may provide a means to distinguish between CDM and SIDM with a constant cross-section $1$ cm$^2$/g at the scale of the Milky Way, if at least $\sim 70$ massive black hole mergers are seen over the four-year mission, though more work is needed to determine this with certainty. This value ($\sim$70 mergers) should not be regarded as a quantitative requirement for LISA, given the limitations of our small sample size. Cosmic variance may considerably change the number of mergers required, which could be evaluated in future work incorporating many more simulations.

Due to our limited sample, our results should be seen as illustrative instead of definitive, though they provide an intriguing hint that LISA could provide new insights into dark sector physics. As the uncertainty in the number of massive black hole mergers LISA will observe is large, $\sim 70$ mergers may or may not be observed. LISA might see only a few such mergers per year, or it might see a hundred per year~\citep{LISA_Detection_Rate_and_Review}. Despite limitations, this work is exciting because it offers a hint that LISA may be able to see a difference between CDM and SIDM-1 (SIDM with a constant cross section of 1 $\rm{cm^2/g}$). 

Future work could involve a far greater number of simulations with a variety of halo masses. Our results depend on the SIDM cross section used, so future work could vary the cross section of the SIDM particles, or use a velocity-dependent cross section, which would be more astrophysically relevant. It should be noted that the simulations assume that black holes merge when they are still separated by several hundred parsecs, though in reality, the black holes must become bound as a binary whose orbital separation must then shrink by about a factor of a million before they really merge, a complex process which depends on the stellar and gas content of the galaxy, as well as its central structure, morphology, and kinematics -- all of which are unresolved in these simulations. Though real black hole mergers are certainly delayed, it is not immediately clear how this would affect the relative difference between black hole mergers in SIDM-1 and CDM; for example, in SIDM-1, the dark matter density profile is flattened while the stellar density is increased -- this in principle could argue that including stellar scattering in the merger prescription would reduce the difference between SIDM-1 and CDM. However, early on, the massive black holes themselves are less massive in SIDM-1 than in CDM, which would lengthen the dynamical friction timescale over the last few hundred parsecs, and a putative dark matter spike in SIDM-1 could well hasten the merger in the last parsec~\citep{SIDM_Final_Parsec}. It is beyond the scope of this paper to include all these non-linear effects; future work should focus on incorporating more accurate black hole dynamics within simulations.

In our simulations, massive black hole seeds form once gas particles have a density of 45 $\rm{cm^{-3}}$, a temperature between 9,500 and 10,000 Kelvin, and a metallicity of $Z < 3 \times 10^{-4}$~\citep{Sanchez_2019, 13}. Altering these assumptions, such as the density threshold, could cause seeds to form at different times, thereby shifting when mergers would be observed. This could work to delay mergers by delaying seed formation, though since this would occur for both CDM and SIDM-1, it would not necessarily wash out the delaying effect of SIDM-1 relative to CDM. This highlights another important direction for future work. We also note that the resolution in the simulations was too low to form cores through baryonic feedback, which suggests that future work should involve higher-resolution simulations. Baryonic feedback, when its effects are more accurately resolved, could influence the potential that the black holes sit in, possibly resulting in some additional delay to mergers, separate from the effects of SIDM.    

\section{Acknowledgements}

NNS acknowledges support from the NSF MPS-Ascend award ID 2212959. KHB acknowledges the support of NSF 2125764. The simulations run in this work used resources provided by the NASA High-End Computing (HEC) Program through the NASA Advanced Supercomputing (NAS) Division at Ames Research Center.

\bibliography{Probing_Nature_of_Dark_Matter_Self_Interactions}
\bibliographystyle{aasjournalv7}



\end{document}